\documentclass[12pt]{article}

\usepackage{graphicx}
\usepackage{subfigure}

\newfont{\feff}{cmti10}
\begin{document}

\title{A Cross-Over in the Enstrophy Decay in Two -Dimensional   Turbulence in a Finite Box}
\author{ Victor Yakhot, John Wanderer\\
Department of Aerospace and Mechanical Engineering,\\ Boston
University, Boston 02215}

\maketitle \begin{abstract}
\noindent  The numerical simulation of two-dimensional decaying 
turbulence in a large but finite box presented in this 
paper uncovered two physically different regimes of enstrophy decay.
During the initial stage, the enstrophy $\Omega$, generated by a random Gaussian initial condition, decays as
$\Omega(t)\propto t^{-\gamma}$  with $\gamma\approx 0.7-0.8$. After that,
the flow undergoes a transition to a gas or fluid composed  of distinct  vortices. 
Simultaneously, the  magnitude of  the decay 
exponent crosses over to $\gamma\approx 0.4$. An exact relation for the total
number of vortices, $N(t)$, 
in terms of the mean circulation $\Gamma$ of an individual vortex 
$N(t)\propto 1/\overline{\Gamma^{2}}$  is derived.
A theory predicting  $N(t)\propto t^{-\xi}$ 
and the magnitudes of exponents $\gamma=2/5$ and $\xi=4/5$ is presented and 
the possibility of an additional very late-time cross-over to $\gamma=1/3$ and $\xi=2/3$ is also discussed.

\end{abstract}
\newpage

\noindent \noindent The interest in the problem of two-dimensional (2D) turbulence 
was excited by the pioneering work of
Onzager$^{1}$, who recognized the role of point-vortices in 2D turbulence dynamics and
suggested a theory based on some general 
concepts of statistical physics. 
In Onzager's days the  quantitative investigation of  
two-dimensional hydrodynamics was inhibited by the pausity
of well-controlled physical experiments. Now, with advent of powerful
computers, various  features  of the problem are much 
better understood due to the remarkable works of McWilliams and collaborators$^{2-5}$, 
Benzi et. al.$^{6}$ 
and many others.  Still, many important questions remained unanswered.

If a fluid in a large three-dimensional vessel of linear dimension $L$ is stirred 
on a length-scale ${\cal L}\ll L$, the non-linear process of  generation of the 
small-scale velocity fluctuations leads to energy dissipation,  making a 
steady state possible even 
in the limit of vanishing viscosity. In this case, universality means 
 independence of the 
small-scale features of the flow upon length-scale $L$. In two dimensions, the situation is very different; the Euler equations conserve both energy and enstrophy.  As a result, the most prominent 
dynamic feature of 2D-turbulence are two 
fluxes in the wave-number space$^{7}$. The (negative) energy flux leads to formation of the slow-varying (low-frequency) large-scale motions containing almost the entire energy of the system on a time-dependent integral
scale $l_{0}(t)$. The second, enstrophy flux, leads to generation of 
rapidly-varying  small-scale velocity fluctuations responsible for the enstrophy of the flow.  The short-time dynamics, when $l_{0}(t)/L\ll 1$, are
 characterized by a quasi-steady  inertial range with the energy spectrum
$E(k)\propto k^{-\frac{5}{3}}$ and the integral scale 
$l_{0}(t)\approx 2\pi/k_{0}(t)\propto t^{\frac{3}{2}}\ll L$.
The most interesting phenomenon happens later in the evolution
when $l_{0}(t)\approx L$; the energy spectrum starts piling up at $k\approx \pi/L$ forming very slow varying and very powerful motions where the major fraction of kinetic energy is concentrated$^{8-13}$. 
The enstrophy associated with these motions is negligibly small. Simultaneously, powerful 
small-scale vortices, containing almost the entire enstrophy of the flow, 
are formed at the centers of these large scale structures. The characteristic
scale (radius),$a(t)\approx 1/k_{f}\approx const$, of these vortices is very small, $a(t)\ll L$.
These complex structures depend on geometry of the problem and thus are not
universal$^{8,12}$.

\noindent The time evolution of decaying two-dimensional turbulence also has three distinct intervals. 
During the initial stage,  the "inverse cascade" leads to population of the scales 
$k<k_{0}(t=0)$ with formation of  vortices which merge and create the ever larger 
vortices. At the same time the enstrophy dissipation and some intermittency 
are generated around small-scales$^{14}$. The total kinetic energy in the system remains almost constant while the vortex merger persists until only two coherent vortices with positive and negative vorticity  are formed (McWilliams$^{2}$). 
 
\noindent Depending on the setup, the experimental and numerical data on  the enstrophy decay vary rather dramatically. It has been shown that after some transient, the enstrophy $\Omega$, in turbulence initially generated as a Gaussian 
random velocity field, decays  with time as
$\Omega(t)\approx t^{-\gamma}$ with $\gamma\approx 0.7-0.8^{15}$.  
The recent theory$^{16}$  of 2D-turbulence decay in an infinite system led to an asymptotic magnitude 
$\gamma=2/3$, which is close to numerical results.
Later in the evolution, after a substantial depletion of the initial 
enstrophy, formation of the   
well-separated vortices containing  almost entire enstrophy of a flow has been
observed$^{2}$. This phenomenon inspired various investigators to study the dynamics of 2D-turbulence as evolution of a set of initially created point-vortices$^{3-6}$.  This set up, however, yielded  a different decay law for $\Omega(t)$ with $\gamma\approx 0.35-0.4^{2-6}$.

\noindent To explain this magnitude of the decay exponent, Carnevale et.al.$^{3}$ based their scaling theory on numerical results by McWilliams$^{2}$ giving the total number of vortices in a flow $N\approx t^{-\xi}$ 
with $\xi\approx 0.7-0.75$. More recent simulations by Bracco et al$^{4}$
confirmed this result giving $\xi=0.75\pm 0.05$. Observing  that the
 entire kinetic energy, $K$, of the system is concentrated 
in a rotational
motion of the well-separated vortices, the authors of Ref.[3] obtained 
$K\approx \rho N(t)\omega^{2}_{o}a^{4}=const$ and assuming that the   
peak 
vorticity of 
a  vortex $\omega_{o}\approx const$, 
they obtained  their  decay law: $a(t)\propto N^{-\frac{1}{4}}$ 
$\Omega(t)\approx N(t)\omega_{o}^{2}a^{2}(t)\propto t^{-\frac{\xi}{2}}$ 
in a good agreement with  numerical data.
The observed value of the exponent $\xi\approx 0.75$ remained unexplained. 
The theory of Benzi et. al.$^{6}$
was based on similar assumptions and led to almost qualitatively 
identical results. Unification of the outcomes obtained from these different setups is one of the goals of the work described below.

\noindent In this paper we report the results of a numerical investigation of the 
2D turbulence decay in a large but finite square. (The definition of "large" is given below.) 
It is shown that after an initial enstrophy decay, characterized by 
$\gamma\approx 0.7-0.8$, 
the flow undergoes transition to a state that  can
be represented as a gas of vortices. The enstrophy decay law in this state is 
$\Omega\propto t^{-\gamma}$ with $\gamma\approx 0.4$. 
The theory developed in this paper leads  
to the characteristic length-scale  of a vortex  
$a(t)\propto N^{-\frac{1}{4}}$ exactly as in Ref.[3]. 
The proposed analytic model for the vortex interaction  
gives for the number of vortices in a system as $N(t)\propto t^{-0.8}$ and 
$\Omega\propto t^{-\frac{2}{5}}$, and are in a close agreement with numerical simulations. 
An exact expression for $N(t)\propto 1/\overline{ \Gamma^{2}}$, where 
$\Gamma$ is a circulation of an individual vortex,  
is derived directly from the Navier-Stokes 
equations.\\

The problem of decay of two-dimensional (2D) turbulence we are interested in this paper is formulated in a following way.
Consider a time evolution of an initial velocity field, $u(x,y,0)={\bf u}_{0}$. defined on a two-dimensional square such that $-L\leq x,y\leq L$. 
The field $u_{0}$ is a Gaussian random noise defined in the Fourier space in the vicinity 
of $k=k_{0}(t=0)=O(\frac{N\pi}{L})$ with $\overline{{\bf u}^{2}_{0}}=O(1)$ and $N=const=O(1)$.
At the initial instant, $t=0$, the enstrophy is $\Omega=\overline{\omega^{2}}=\overline{(\nabla \times {\bf u})
^{2} }=O(1)$. The square is large meaning that we are interested in the limit $\nu\rightarrow 0$ and 
$k_{0}(t=0)L\rightarrow\infty$. Still, the box is finite so we will be able to
study both short time,
when $k_{0}(t)L\gg 1$, and  
long time asymptotics when $k_{0}(t)L\approx 1$. In all our simulations the initial kinetic energy of the flow
$K(t=0)=\frac{1}{2}\rho \int v^{2}({\bf x},t=0)d{\bf x}=\frac{1}{2}$ and the fluid density $\rho=1$.

\noindent 
The Navier-Stokes equations with the $O(\nu_4 \nabla^{4}{\bf v})$ hyper-viscous dissipation  terms were simulated 
using a pseudo-spectral method. To be sure that hyper-viscosity did not influence the results,
two simulations with resolutions $1024^{2}$ and $2048^{2}$ were conducted. The
initial random field was Gaussian with energy spectrum 
$E(k)=a_s \frac{K}{\rho  k_p} \left(\frac{k}{k_p}\right)^{2 s +1} e^{-\left( s+ \frac{1}{2}\right)\left(\frac{k}{k_p} \right)^2 } $ where $a_s=\frac{(2s+1)^{(s+1)}}{2^s s!}$ and $s=3$ as in Ref.[15].  The parameters of the simulations are given in the following table.

\begin{table}[h]
\center
\begin{tabular}{|c|c|c|c|c|c|}
\hline
 N & time-step & $\frac{2K}{\rho}$ & $\Omega(t=0)$ & $k_p$ & $\nu_4$ \\
\hline
\hline
 $1024$ &$ 0.0005$ &$ 1$ &$ 1170$ &$ 32$ & $2.1x10^{-10} $ \\
\hline
 $2048$ & $0.00025$ &$ 1$ &$ 1170$ &$ 32$ & $9.6x10^{-11} $ \\
\hline
 \end{tabular}
 \caption{Numerical simulation parameters}
  \label{fig:num_params}

\end{table}

\noindent On Fig.1 we show the time evolution of both enstrophy, $\Omega$, and energy, $K$, in a flow. 
 During the entire simulation, kinetic energy stays constant; this is not so for the enstrophy. 
Initially after some transient, the enstrophy $\Omega(t)$ obeys the power law $\Omega\propto t^{-\gamma}$ 
with $\gamma\approx 0.7-0.8$ and is very close to the result reported before by Chasnov$^{15}$. 
\noindent Later in the evolution, when the dimensionless time $T$ is such that 
$T=\sqrt{\overline{\omega^{2}}(t=0)} t\geq 200$,  a cross-over to the value 
$\gamma\approx 0.4$ is clearly seen. On Figs. 2-3 the time evolution of vorticity field is presented. 
We can observe that the cross-over to a new long-time decay regime ($T\geq 200$), with $\gamma\approx 0.4$, is accompanied by formation of the well-separated distinct vortices.  
It is interesting, that similar structures were observed in case of the small-scale forced 2D-turbulence$^{9,10}$.
One feature of Figs. 2-3 deserves a special attention. At the initial
 stage of decay, $10 \leq T \leq 200$, the vortices are more or less homogeneously packed, while at the later times the vortex density 
shows a clear tendency of the flow structures to clamp together keeping the mean distance between the 
neighbors 
more or less constant.  This may be an indication of a weakening of the screening of the inter-vortex interaction as the vortex density decreases.
The merger itself happens in two stages; first the same-sign vortices approach
each other with 
translational velocity $v_{o}$ 
and start moving as a couple ("vortex molecule") 
rotating around their mutual center of mass. The merging process is relatively slow 
meaning that during this "pre-merger" period, the characteristic frequency of this joint motion is $\omega_{m}$ where $\omega_{m}\approx v_{o}/a(t)\ll\omega_{o}$ and $\omega_{o}$ is the mean
angular velocity of an individual vortex. 
This "dance" is interrupted by a collapse with a single emerging vortex with the peak vorticity $\omega_{o}\approx constant$.\\

\begin{figure}[h]
  \center
  \subfigure[($1024^2$)]{\includegraphics[height=6cm]{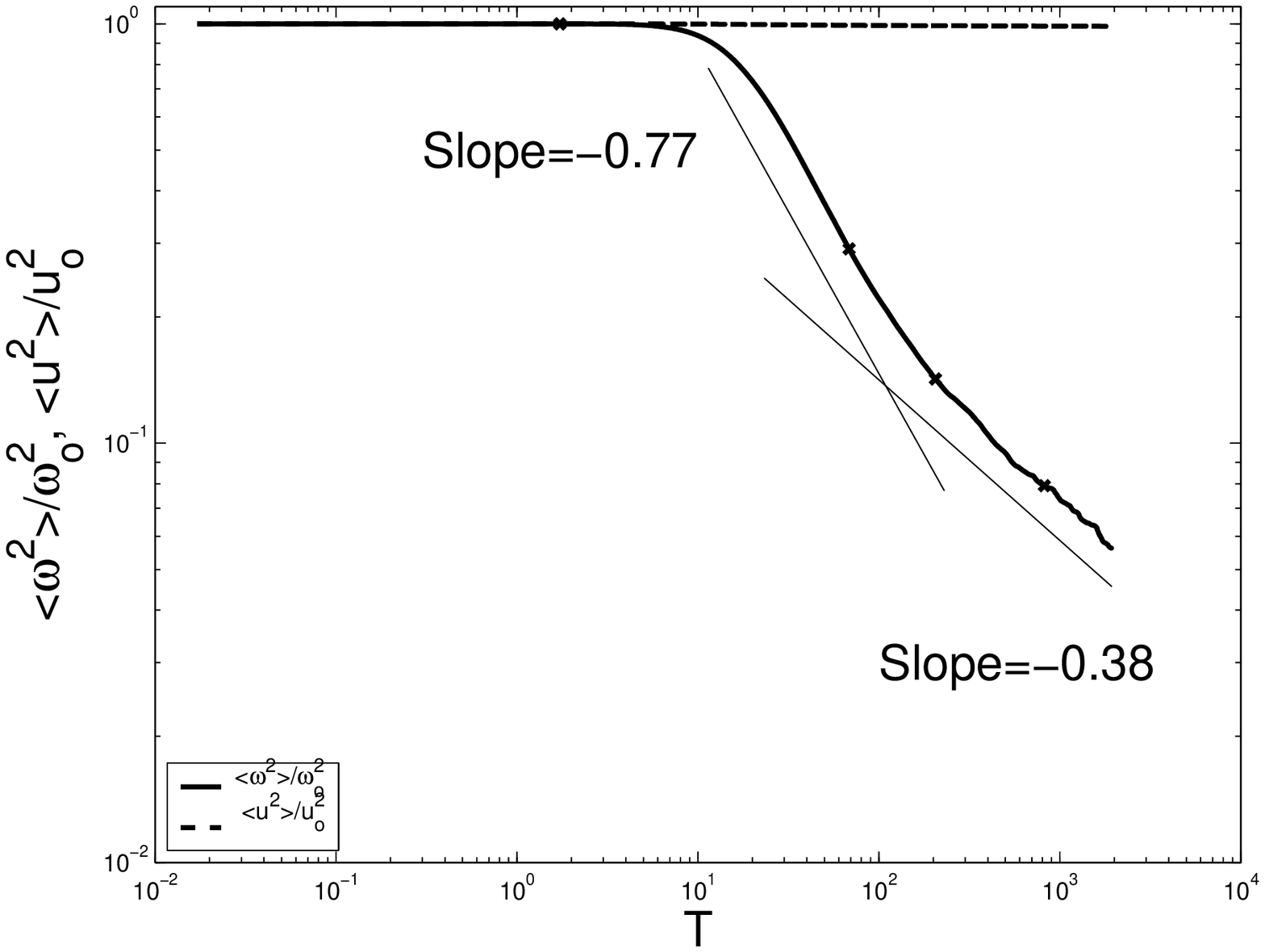}}
  \subfigure[($2048^2$)]{\includegraphics[height=6cm]{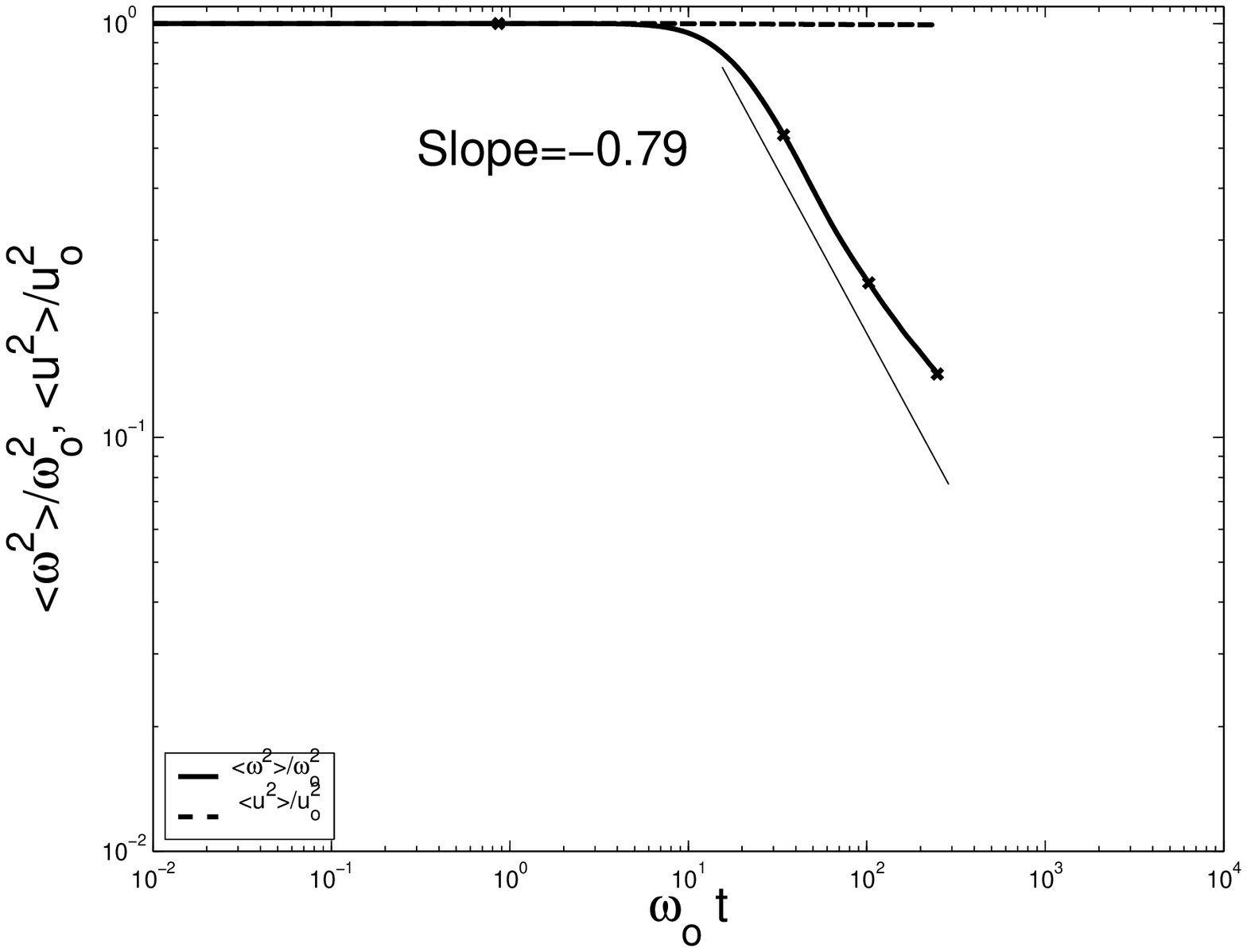}}
  \caption{Enstrophy and energy decay as function of time }
  \label{fig:enstrophy_decay}
\end{figure}

\begin{figure}[h]
  \center
  \includegraphics[height=6cm]{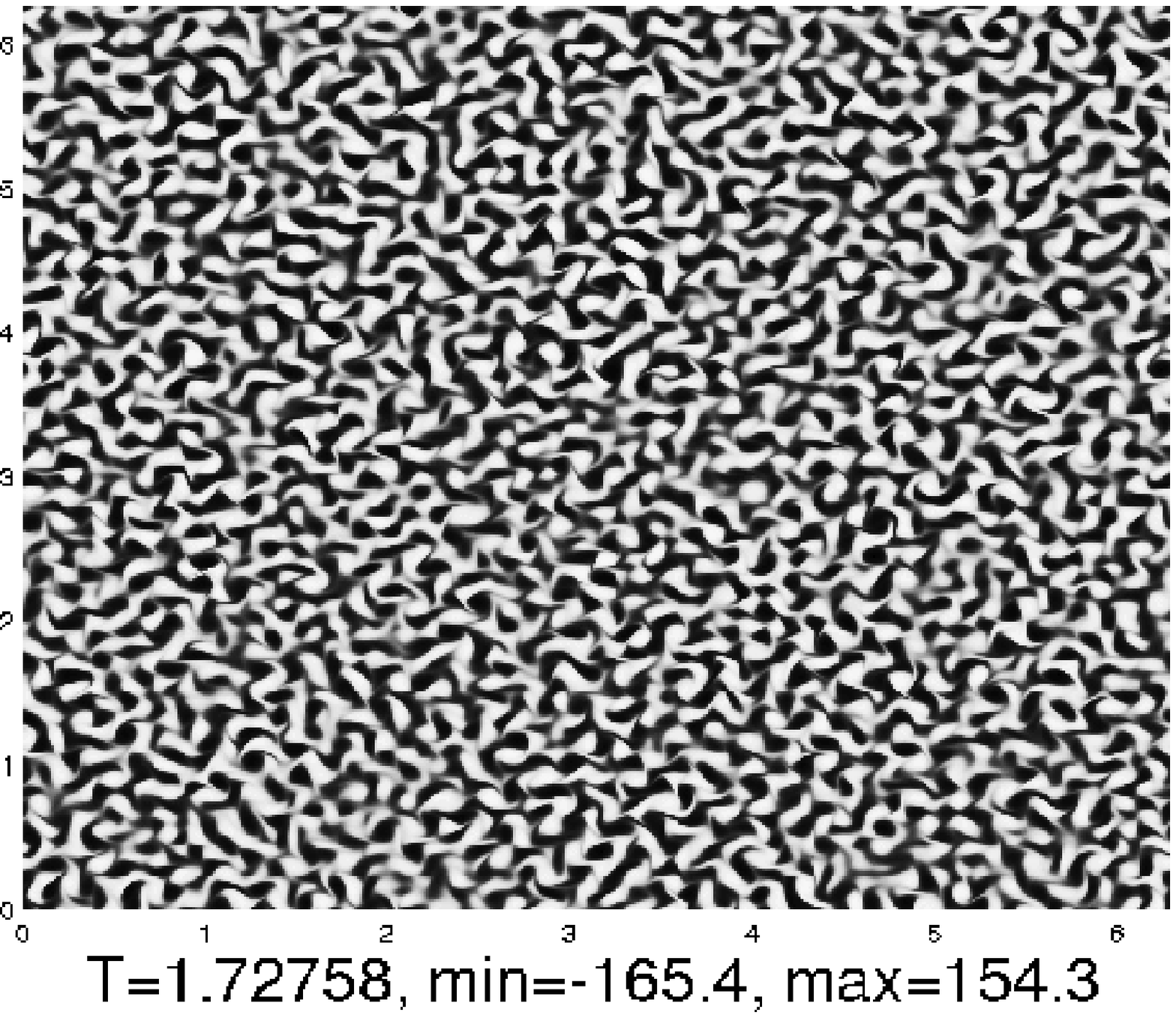}
  \includegraphics[height=6cm]{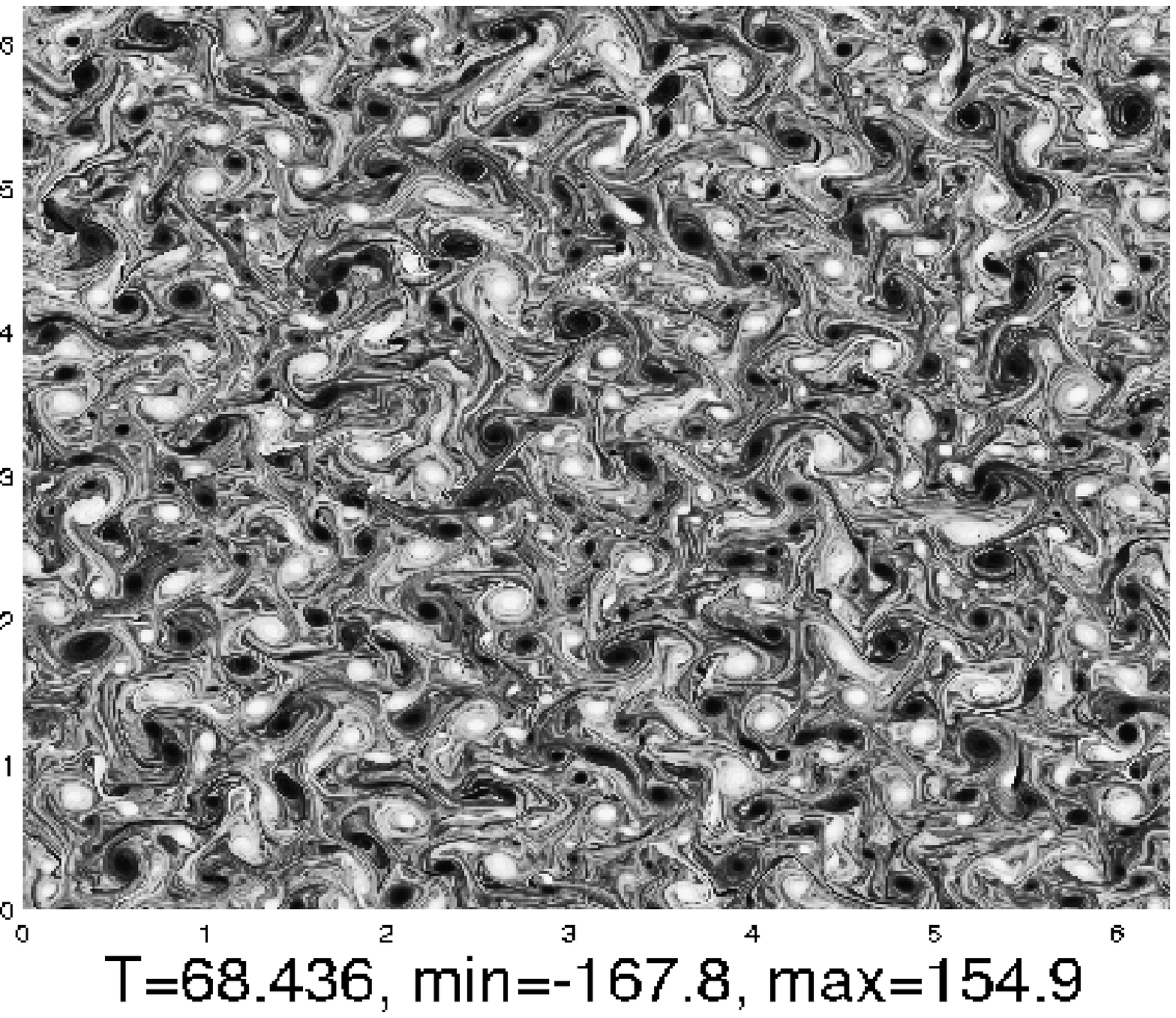}
  \caption{Vorticity field before crossover ($1024^2$)}
  \label{fig:omega_1024_early}
\end{figure}

\begin{figure}[h]
  \center
  \includegraphics[height=6cm]{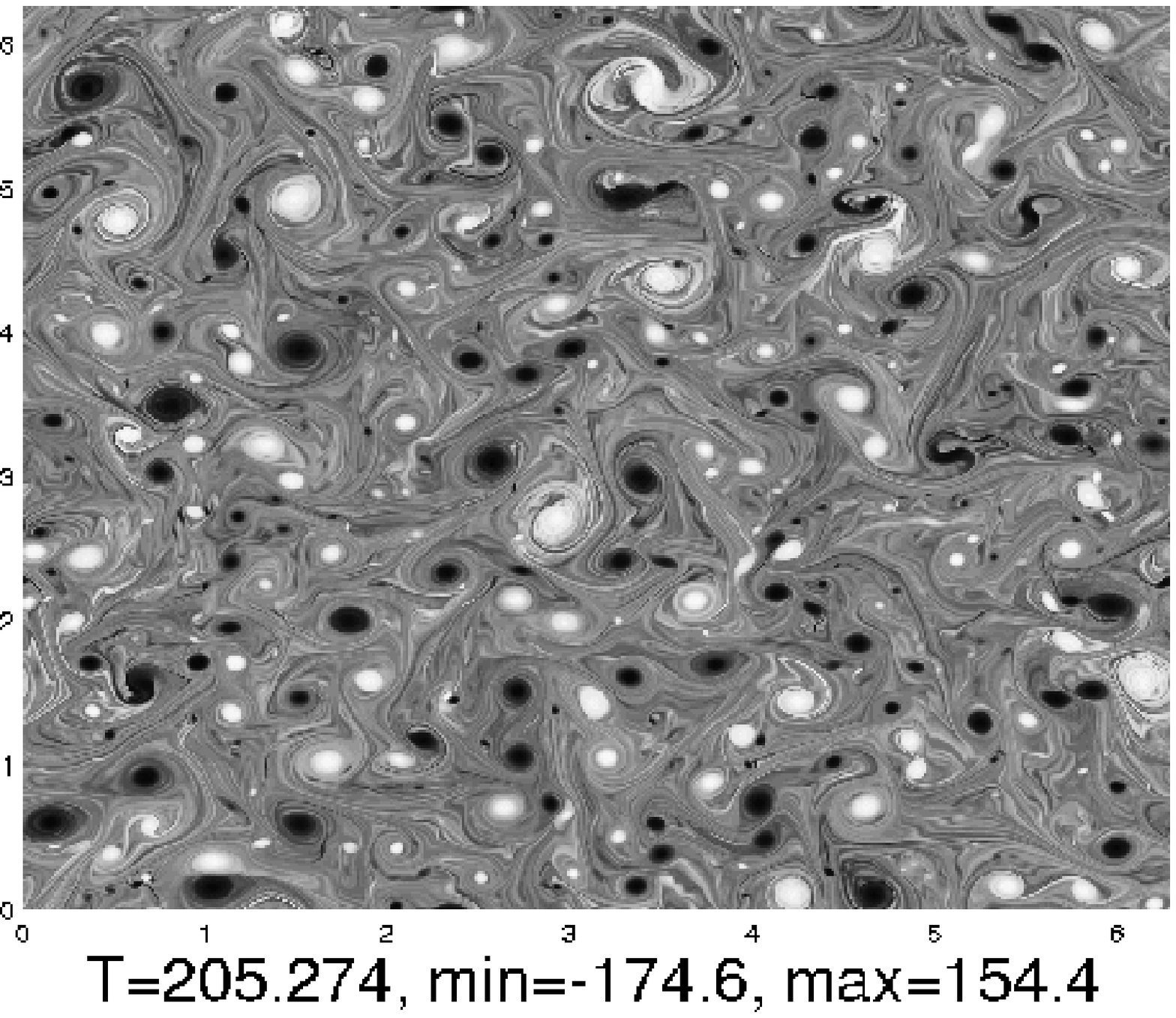}
  \includegraphics[height=6cm]{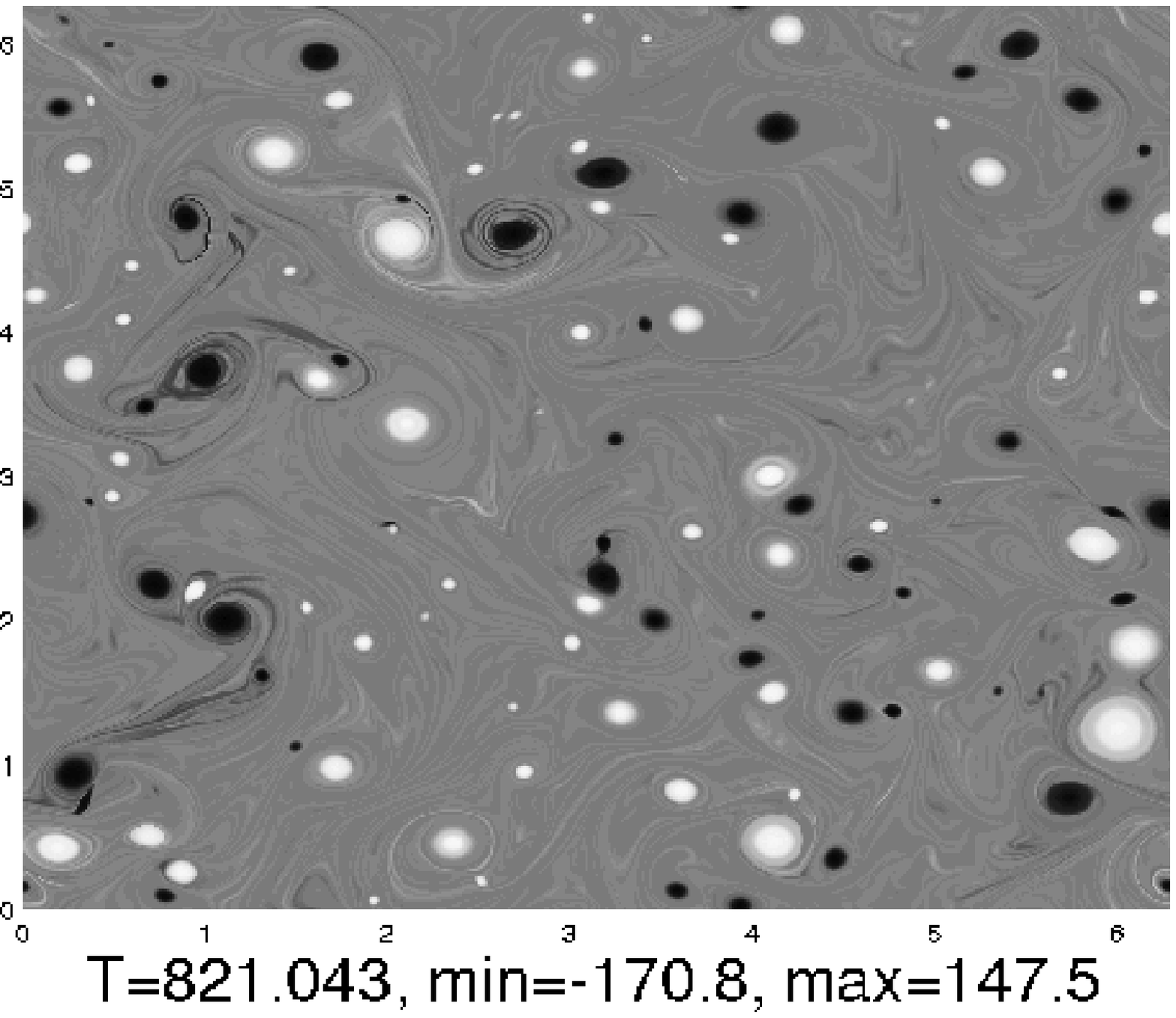}
  \caption{Vorticity field during and after crossover ($1024^2$)}
  \label{fig:omega_1024_late}
\end{figure}

\begin{figure}[h]
  \center
  \includegraphics[height=6cm]{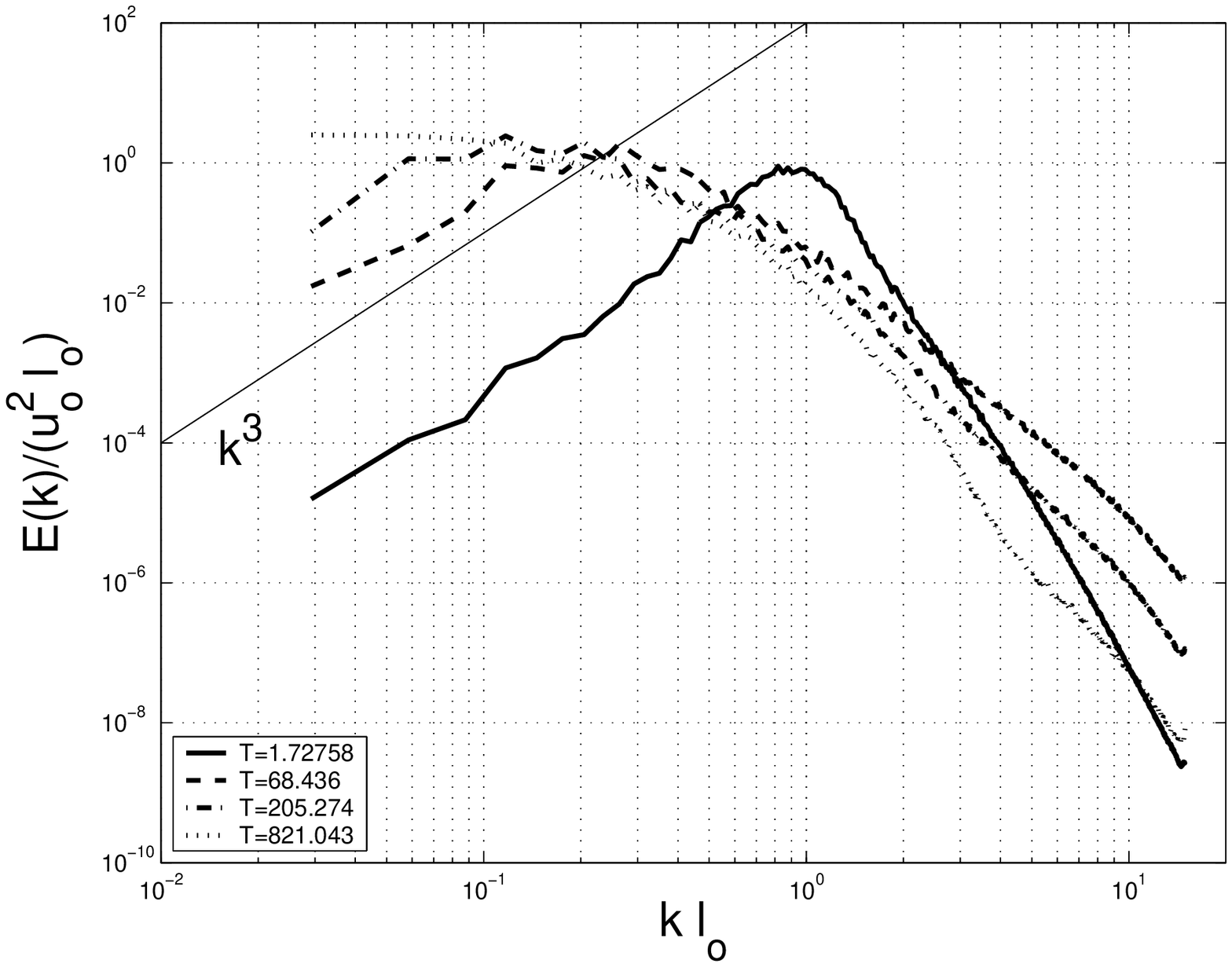}
  \caption{Time evolution of energy spectrum ($1024^2$)}
  \label{fig:energy_spectrum}
\end{figure}

\noindent First we present a model which is a generalization of a remarkable 
work by Carnevale et.all.$^{3}$.
After that we will demonstrate its consistency with the Navier-Stokes equations.
The total kinetic energy of a flow ($[K]=grcm^{2}/sec^{2}$) is: 

\begin{equation}
K(t)=\frac{\rho}{2}\int v^{2}({\bf x})d^{2}x =\frac{1}{2}\rho N(t)[v_{o}^{2}a^{2}+\omega_{o}^{2}a^{4}]=K(t=0)=O(1)=constant
\end{equation}
\noindent  
We conjecture and later verify that while the $O(v_{o}^{2}a^{2})$ contribution to the total energy is small, 
it entirely dominates the time-scale of the merger process. According to Carnevale et al$^{3}$ 
 and 
our own numerical simulations we assume that $\omega_{o}\approx const$. 
Neglecting the $O(v_{o}^{2})$ , the equation (1) leads to the estimate 
obtained in Ref.[3]: 
$\frac{1}{2} \rho N(t)\omega_{o}^{2}(t)a(t)^{4}=K=const$
and  $a(t)\approx (\frac{2K}{\rho N(t)\omega_{o}^{2}})^{\frac{1}{4}}\propto N^{-\frac{1}{4}}$

\noindent Following Carnevale et al$^{3}$, we introduce the mean distance
between the nearest- neighbor vortices as 
$R\approx L/\sqrt{N(t)}$ and estimate translational velocity $v_{o}$ of a vortex
moving in the field of other vortices: as:

\begin{equation}
v_{o}\approx \frac{\Gamma}{R(t)}\approx
\frac{\omega_{o}a^{2}\sqrt{N}}{L}=\sqrt{\frac{2K}{L^{2}\rho}}\equiv V
\end{equation}
\noindent where $\Gamma\approx \omega_{o}a^{2}$ is a circulation. 
To complete the argument, we estimate the rotational velocity of a vortex as: 
$v_{R}\approx \omega_{o}a(t)
=(\frac{2K\omega_{o}^{2}}{\rho N(t)})^{\frac{1}{4}}$. Thus, 
the ratio
$\frac{v_{o}}{v_{R}}=
(\frac{V^{2}}{\omega_{o}^{2}L^{2}})^{\frac{1}{4}}N^{\frac{1}{4}}\propto 
t^{-\frac{\xi}{4}}\rightarrow 0$ and  
is indeed small. 
It is important that according to the above estimate, while $v_{R}$ grows with time, 
translational velocity $v_{o}$ is a time-independent constant.

\noindent Now we need a model for $N(t)$. We consider a vortex merger process 
as a binary ``chemical reaction'' 
$2N\Longrightarrow N$ with the "reaction rate" ${\cal R}(t)$, where is the inverse of the time the vortex pair spends in a bound state prior to the merger event.
If vortices are well separated, the population equation is:

\begin{equation}
\frac{dN}{dt}=-{\cal R}N^{2}(t)
\end{equation}

\noindent 
We recall  that two same-sign  vortices attract each other,  tending to form 
a spinning dipole (bound state) with the linear dimensions (radius of the
orbit) depending upon the impact parameter, 
relative translational velocity, and other characteristics of the collision
process. It is important that the larger the relative  velocity , the
larger the radius of an orbit of a bound  state. 
The rotational frequency of this bound pair is $O(v_{o}/I)$ where $I$ is the
impact parameter. 
In the system we are dealing with, the binary collisions happen
 in the field of other 
 vortices of both signs often strongly perturbing the pre-merging dipole. 
It is clear, and we observe it in
our simulations,  that only the most strongly interacting pairs  stay stable
under perturbations from  surrounding fluid long enough to eventually merge. 
The most stable ones are those
with  linear dimension (impact parameter) not larger than a few $a(t)$ and
having smallest relative velocity. 
 Since the translational velocity $v_{o}\approx const$,
 describing the motion of coupled vortex pair before the merger event, 
 is the smallest velocity-scale in the system we choose it as a rate determining parameter. 
The merger rate is thus estimated as: 
${\cal R}\approx v_{o}/a(t)\approx v_{o}(\frac{N(t)\omega_{o}^{2}\rho}{2K})^{\frac{1}{4}}=
\frac{v_{o}}{L}\sqrt{\frac{\omega_{o}L}{v_{o}}}N^{\frac{1}{4}}$. 
Substituting this into the above differential 
equation gives:

\begin{equation}
N=\frac{N_{0}}{(\frac{t}{t_{0}}+1)^{\frac{4}{5}}}\rightarrow N_{0}(\frac{t}{t_{0}})^{-\frac{4}{5}}
\end{equation}

\noindent where $t_{0}\approx \sqrt{\frac{L}{v_{o}\omega_{o}}}$ and $N_{0}$ is an undetermined constant. This gives the enstrophy 
decay law as $\Omega(t)\approx N(t)\omega_{o}^{2}a^{2}(t)\propto t^{-\frac{2}{5}}$ and is 
extremely close to  numerical result. It follows from (4) 
that in an infinite system ($L\rightarrow \infty$)  
the cross-over to $\gamma\approx 0.4$
never happens and   the enstrophy 
decay process with $\gamma=2/3$ persists in the limit $t\rightarrow \infty$.
Now we would like to discuss another possibility.
If the mean distance between the nearest vortices is large ($R=O(L/\sqrt{N(t)}$), then the reaction rate
${\cal R}=O(v_{o}\sqrt{N}/L)$ and equation (5) gives $N(t)\propto t^{-\frac{2}{3}}$ and 
$\Omega(t)\propto t^{-\frac{1}{3}}$. \\
 
\noindent The relations derived above are the outcome of a kinetic-like considerations. 
It is interesting to verify their consistency with the Navier-Stokes equations.
 Choosing the displacement vector ${\bf r}$ parallel to the $x$-axis and assuming statistical 
 homogeneity and isotropy of a flow , the Navier-Stokes equations lead to the relation for the correlation 
function 
of the scalar vorticity $D_{\omega}=\overline{\omega(x)\omega(x+r)}^{16-19}$:

\begin{equation}
\frac{\partial D_{\omega}}{\partial t}=-
\frac{1}{4r}\frac{\partial}{\partial r}r
\overline{(v(x+r)-v(x))(\omega(x+r)-\omega(x))^{2}}
\end{equation}

\noindent 
Multiplying this equation by $r$ and integrating over  the interval $0\leq r \leq L$  gives$^{16-18}$:
\begin{equation}
\frac{\partial}{\partial t}\int dr r\overline{\omega(x+r)\omega(x)}=-
\frac{1}{4}r\overline{(v(x+r)-v(x))(\omega(x+r)-\omega(x))^{2}}|^{L}_{0}
\end{equation}

\noindent In the limit $L\rightarrow \infty$ the right side of this equation is equal to zero. 
To evaluate the integral, we recall that in 2D hydrodynamics  the point-vortex
representation of vorticity is: $\omega({\bf x})=\sum_{i}\Gamma_{i}\delta({\bf
  x}-{\bf X}_{i}(t))$, where ${\bf X}_{i}(t)$ is a position of the $i^{th}$ vortex and $\Gamma_{i}$ stands for its circulation.
Then, recalling that the flow is filled with vortices of both signs so that  $\sum_{i\neq j}\Gamma_{i}\Gamma_{j}=0$,
we derive readily:
\begin{eqnarray*}
\int dr~r\overline{\omega({\bf x})\omega({\bf x}+r{\bf i})}=
\int d{\bf r}\frac{1}{2\pi L^{2}}\sum_{i,j}\Gamma_{i}\Gamma_{j}
\int d{\bf x}\delta({\bf x-X_{i}(t)})\delta({\bf x -X_{j}(t)}+r{\bf i})=\\
\sum_{i,j}\frac{\Gamma_{i}\Gamma_{j}}{2\pi L^{2}}=\sum_{i=1}^{N(t)}\frac{\Gamma_{i}^{2}}{2\pi L^{2}}=
\frac{N(t)\overline {\Gamma^{2}}}{2\pi L^{2}}\approx\frac{N(t)\omega_{o}^{2}a^{4}}{2\pi L^{2}}\approx const~~~~~~~~~~~~~~~~~~~~~~~~~~~~~~~~~~~~~~~(7)
\end{eqnarray*}

\noindent This  relation, derived from purely hydrodynamic considerations, is quite important since it 
serves as a correct definition of both $\omega_{o}$ and 
$a(t)$ which are the basic ingredients of the "kinetic theory" developed  in Carnevale et.al$^{3}$. The average- splitting  
$\overline{\Gamma^{2}}\approx \omega_{o}^{2}a^{4}$, used in derivation of the last relation in (7) is an 
approximation valid only if the peak vorticity and  vortex radius are statistically independent. 
At the present time we cannot assess 
the quality of this approximation.\\
 
We  can also start with the Navier-Stokes equations 
and  write equation for the correlation function of the 
$x$-component of velocity field ( here denoted as $v$)$^{16-18}$:

$$
\frac{\partial}{\partial t}\overline{v(x+r)v(x)}=
\frac{1}{6r^{3}}\frac{\partial}{\partial r}r
\overline{(v(x+r)-v(x))^{3}}\eqno{(8)}
$$

\noindent Repeating the considerations leading to (6) we obtain:
$\int r^{3}\overline{v({\bf x})v({\bf x}+{\bf i}r)}dr=const$ and  
introducing the correlation length $a(t)$, this integral is estimated readily
giving
$
N(t)v^{2}a^{2}\approx N(t)(v_{o}^{2}+\omega_{o}^{2}a^{4})\approx const
$
where $N(t)=L^{2}/a^{2}$. This is exactly the relation (1) evaluated above from the energy conservation 
considerations. \\

\noindent The main result of this work is unification of the two different exponents $\gamma\approx 0.7-0.8$ and $\gamma\approx 0.4$ as corresponding to intermediate and long time regimes of the enstrophy decay in 
2D turbulence, respectively. The exact relation $N(t)\propto 1/\overline{\Gamma^{2}}$ derived from the Navier-Stokes equation, 
is a direct generalization of an approximate formula $N(t)\propto a^{-4}$ 
obtained  by Carenevale et.al.$^{3}$ using  the  "kinetic theory"  
approach  and assuming 
the constancy of the  peak vorticity of  individual vortices. 

\noindent We have identified three possible values of  exponent $\gamma$. According to ref.[16],
the  initial stage of enstrophy 
decay in a large box ($\L\rightarrow\infty$) is characterized by $\gamma=2/3$ 
which is close to  results of numerical simulations by Chasnov$^{15}$ and the ones presented in this paper. 
It has been demonstrated that in a finite box, after the depletion of the initial externally  
introduced enstrophy, a flow composed of long-living distinct vortices emerges and the exponent crosses over to $\gamma\approx 0.4$,   
very close our theoretical prediction. Another cross-over to $\gamma\approx 1/3$ and $\xi=2/3$  as a very long-time asymptotics 
in a system with small number of vortices is predicted. To the best of our knowledge, this regime has not yet been observed.

\noindent The nature of the transition remains poorly understood. It has been shown$^{9-10}$ that in case 
of the small-scale 
forced 2D turbulence, the well-separated vortices emerge as a result of the finite-size 
effect (condensation). It is hard to say if this conclusion holds for freely decaying 2D turbulence but if it does,
then the question of universality of exponents $\gamma$ and $\xi$ becomes
extremely interesting and important. Some indications that the finite-size
effects can play an important role can be found from Figs.1-4. It can be shown
theoretically that as long as $k_{0}(t)L\gg  1$, the large-scale 
energy spectrum in this developing flow
is $E(k\ll k_{0}(t))\propto k^{3}\phi(t)$. We can see that in the short-time
interval, $T<200$, this is indeed approximately correct and the exponent is
$\gamma\approx 0.7-0.8$. At the longer times ($T>200$),  when
$k_{0}(t)L\approx 1$,  the energy spectrum starts strongly deviate from the 
$k^{3}$-curve. This is definitely due to the finite size- effects. 
Simultaneously,
the well-separated vortices emerge and the exponent crosses over to
$\gamma\approx 0.4$.

\noindent{\bf \Large References.}

\noindent 1.~L. Onzager, Nuovo Cimento {\bf
  6}(2),
279 (1949)\\
\noindent 2~. J.C. McWilliams, J.Fluid Mech {\bf 219}, 361 (1990)\\
\noindent 3.~G.F.Carnevale, J.C.McWilliams, Y.Pomeau, J.B.Weiss and W.R.Young, Phys.Rev.Lett.{\bf 66}, 2735 (1991).\\
\noindent 4.A.Bracco, J.C.McWilliams, G.Murante,A.Provenzale,J.B.Weiss, Phys.Fluids {\bf 12}, 2931 (2000).\\
\noindent 5.~M.V.Melander, J.C.McWilliams, and N.Zabusski, J.Fluid mech. {\bf 178}, 137 (1987).\\ 
\noindent 6~R.Benzi, S.Patarnello and P.Santangelo, J.Phys.A {\bf 21}, 1221 (1988)\\
\noindent 7. R.H.Kraichnan, Phys.Fluids {\bf 10}, 1417 (1967)\\
\noindent 8. U. Frisch and P-L Sulem, Phy.Fluids {\bf 27}, 1921 (1984) \\
\noindent 9. L.M. Smith and V.Yakhot, Phys.Rev.Lett. {\bf 71}, 352 (1993)\\
\noindent 10. L.M. Smith and V. Yakhot, J.Fluid Mech. {\bf 274} (1994)\\
\noindent 11. G.Bofetta, A.Celani and M. Vergassola, Phys. rev.E {\bf 61}, R29 (2000).\\
\noindent 12. J. Paret and P. Tabeling, Phys. Fluids {\bf 12}, 3126 (1998) \\
\noindent 13. R.H.Kraichnan and D. Montgomery, Reports Prog.Phys. {\bf 43}, 547 (1980)\\
\noindent 14. L.M. Smith V. Yakhot, Phys.Rev.E{\bf 55}, 5458 (1997)\\
\noindent 15. J.R.Chasnov, Phys. Fluids {\bf 9}, 171 (1997)\\
\noindent 16.~V.Yakhot, Phys.Rev.Lett. 2004 (in press).\\
\noindent 17. A.S. Monin and A.M.Yaglom, vII, The MIT press, Cambridge, 1975\\
\noindent 18. L.D. Landau and E. M. Lifshitz, Pergamon Press, Oxford, 1982\\
\noindent 19. L.G. Loitsyanskii,Trudy Tsentr. Aero.-Gidrodyn. Inst., {\bf 3}, 33 (1939), (in Russian) \\

\end{document}